\documentclass[runningheads]{llncs}
\usepackage[utf8]{inputenc}

\usepackage{graphicx}
\usepackage{floatrow}

\usepackage{tikz}
\usepackage{pgfplots}

\usepackage[]{algorithm2e}
\usepackage{caption}

\usepackage{makeidx}
\usepackage{amssymb}
\usepackage{amsmath}
\makeatletter
\newcommand\footnoteref[1]{\protected@xdef\@thefnmark{\ref{#1}}\@footnotemark}
\makeatother
\begin{document}
\title{The Potential of Restarts for ProbSAT}
\author{Jan-Hendrik Lorenz \and
	Julian Nickerl}
\authorrunning{J. Lorenz and J. Nickerl}
\institute{Ulm University, Institute of Theoretical Computer Science, 89069 Ulm, Germany 
	\email{\{jan-hendrik.lorenz,julian.nickerl\}@uni-ulm.de}\\
	%\url{http://www.springer.com/gp/computer-science/lncs} 
}
\maketitle              % typeset the header of the contribution

\bibliographystyle{splncs04}

\begin{abstract}
	This work analyses the potential of restarts for probSAT, a quite successful algorithm for k-SAT \cite{balint2014probsat,balint_choosing_2012}, by estimating its runtime distributions on random 3-SAT instances that are close to the phase transition. We estimate an optimal restart time from empirical data, reaching a potential speedup factor of 1.39. Calculating restart times from fitted probability distributions reduces this factor to a maximum of 1.30. A spin-off result is that the Weibull distribution approximates the runtime distribution for over 93\% of the used instances well.
	
	A machine learning pipeline is presented to compute a restart time for a fixed-cutoff strategy to exploit this potential. The main components of the pipeline are a random forest for determining the distribution type and a neural network for the distribution's parameters. ProbSAT performs statistically significantly better than Luby's restart strategy and the policy without restarts when using the presented approach. The structure is particularly advantageous on hard problems. 
\end{abstract}

\section{Introduction}
Las Vegas algorithms often achieve the best performance for many hard problems.
Hence, within these algorithms, many uncertain decisions have to be made. Usually, it cannot be guaranteed that these choices are the best or even good ones. Many bad decisions in a row may lead the algorithm into something like a local optimum that is hard to leave again. 
Once in such a state, it could be beneficial to restart the whole process and retry from the beginning, hoping that this time the algorithm makes better decisions. 

Not all probabilistic algorithms can benefit from restarting, and even if they do in some instances, they might not in general. Additionally, since in many problems no precise information is available on how close the algorithm is to a solution, it is hard to say how long it should wait before restarting. It is known, that if restarts are beneficial for a specific instance, the runtime behavior of the algorithm on that instance can often be described with a probability distribution that possesses a so-called heavy-tail. A distribution is called heavy-tailed if its survival function decays slower than any exponential, i.e.,
the  cumulative distribution function approaches 1 slowly. Gomes et al. \cite{gomes1997heavy} found that a particular kind of heavy-tails, so-called power-laws, describe the runtime behavior 
of combinatorial search procedures very well. They showed that the use of a restart strategy effectively eliminates the heavy-tailed behavior and greatly improves the runtime.
Nowadays restarts are, for example, used in most state-of-the-art SAT solvers \cite{biere2009handbook}. 

Two theoretically intriguing restart strategies are introduced by Luby et al. \cite{luby_optimal_1993}:  The fixed-cutoff approach and Luby's universal strategy. The fixed-cutoff approach minimizes the expected runtime for some fixed restart time. 
However, finding this optimal restart time is a hard task and requires nearly complete knowledge of the algorithm's behavior on the problem instance. 
Luby's universal strategy does not require any domain knowledge. They even showed that there is no other approach without domain knowledge that performs better. Thus, while the fixed-cutoff approach is theoretically the best strategy, Luby's universal strategy is usually preferred (for example \cite{huang2007effect}). Haim and Walsh \cite{haim2009restart} created a portfolio-solver using several restart strategies including the Luby and fixed-cutoff strategy. They apply machine-learning methods to decide which strategy is employed. 

Lorenz \cite{lorenz_runtime_2018} developed a method to calculate the optimal restart time for the fixed-cutoff approach if the runtime behavior
is given by a probability distribution. Hoos and Stützle \cite{hoos1998evaluating} use probability distributions to approximate 
empirically measured runtimes. This is called the (empirical) runtime distribution. Arbelaez et al. \cite{arbelaez2016learning} studied the empirical runtime distribution of probSAT, a stochastic local search SAT-solver by Balint and Schöning \cite{balint_choosing_2012}, and found that lognormal distributions
describe the runtime behavior well. A recent development in the field is trying to predict
the runtime distributions of unseen instances with machine learning methods (see for example Arbelaez et al. \cite{arbelaez2016learning}, Eggensperger et al. \cite{eggensperger_predicting_2017}).

\textbf{Our contribution:}
This work analyzes the capability of probability distributions to describe the runtime behavior of probSAT on random 3-SAT instances close to the phase transition.
We approximate the runtime distributions of probSAT with a similar methodology as presented by Arbelaez et al. \cite{arbelaez2016learning}.
In our work, the generalized Pareto, the lognormal and the Weibull distribution are used to describe the runtime behavior. It is observed that 
especially the Weibull distribution is well suited to describe the runtime behavior of probSAT,
describing significantly more empirical runtime distributions than the established lognormal distribution.

%To evaluate restart times from the fitted distributions, we estimate optimal restart times from the empirical data as a base line. 
We estimate optimal restart times from the empirical data as a baseline to evaluate restart times from the fitted distributions.
Applying this restart time to the observed data leads to an average speedup factor of $1.393$.
The aforementioned fitted distributions are used with Lorenz's method to find theoretically optimal restart times. %The restart times are then evaluated for their potential speedup as compared to not restarting.
We found a potential speedup factor of up to $1.300$ by applying those restart times.
To the best of our knowledge, the potential speedup by restarting at the optimal time according to the respective runtime distributions has not been systematically studied before.

In the second part of this work, a machine learning pipeline which predicts the runtime distribution of so far unseen instances is presented.
It consists of a random forest predicting the distribution type, and a total of three neural networks which predict the parameters of the distributions.
The parameters are used to calculate restart times.  
This approach differs from the procedure introduced by Haim and Walsh \cite{haim2009restart}. The strategies in their portfolio use restart times which are independent of the instance. We observe statistically significant speedups according to both the t-test and a modified Wilcoxon signed-rank test (\cite[Chapter 7.12]{hollander2013nonparametric}). An average speedup factor of more than $1.21$ is observed in hard instances, without worsening the performance on easier ones.
An overall speedup factor of $1.06$ is achieved. 

\section{Restarts and Runtime Distributions}

This section introduces the concept of restarts and the considered runtime distributions.

\subsection{Restarts}
Throughout this work we use the so-called fixed cutoff strategy \cite{luby_optimal_1993}: The algorithm is always restarted after the same number of steps, and
the number of possible restarts is unbounded. It is known that there is an optimal fixed restart time $t^*$ which minimizes the expected runtime. If the restart time $t^*$ is used, then the fixed cutoff strategy is superior to all other restart approaches. However, finding the best possible restart time $t^*$ is a hard task.

\subsection{Runtime Distributions}
\label{sec:bla}
The runtime of a randomized algorithm can be interpreted as a random variable.
The long-term behavior has been extensively studied in survival analysis.
Even though runtimes are discrete values, it is common to model the performance of an algorithm as a continuous process to use the tools provided by survival analysis.

%To use the tools provided by survival analysis it is common to model the performance of an algorithm by
%a continuous process, even though runtimes are discrete values. 
We follow this convention and choose to use the lognormal, the Weibull, and the generalized Pareto (GP) distribution to describe the runtime behavior of probSAT.
All three types of distributions are often used in the fields of survival analysis and reliability engineering. 

The lognormal, the Weibull and the GP distribution were already considered as suitable in previous research articles.
Most notably Arbelaez et al. \cite{arbelaez2016learning} observed that the runtime behavior of randomly generated 3-SAT instances with a clause to variable ratio of $4.2$ can be described by lognormal distributions. Other results favoring the lognormal distribution are given by Arbelaez et al. \cite{arbelaez_using_2013} and Truchet et al. \cite{truchet_estimating_2016}. The former paper found that the runtime distributions (RTDs) of other SAT-solvers, CCSAT and Sparrow, also follow lognormal distributions. The latter observed that the RTD of the CSP-solver Adaptive Search on magic-square-problem instances is well described by the lognormal distribution.

%\begin{figure}[htbp]
%	\input{images/weib_fit}
%	\caption{The empirical distribution function and the Weibull fit of a typical instance.}
%	\label{fig:compare_weib}
%	% weib: k3-n1500-m6361-r4.241-s2879353699
%\end{figure}

Algorithms that show a runtime behavior which resembles a Weibull distribution have also been observed before. Hoos and St\"{u}tzle \cite{hoos1998evaluating} investigated the performance of the GSAT algorithm for different noise parameters. They noted that 
the RTD is approximately a Weibull distribution if the noise parameter is less than the (empirically observed) 
optimal noise parameter. Frost et al. \cite{Frost1997CSP} argued that the RTDs of several CSP-solver can be modeled
by Weibull distributions for satisfiable, binary instances. 

Another often encountered phenomenon in the analysis of algorithms is the occurrence of power-laws. 
Power-law distributions are defined by their tail behavior: If the probability density function behaves like a polynomial, then the distribution follows a power-law.
This property is usually expressed in terms of the Pareto distribution. Such distributions are mainly found in combinatorial
search algorithms. Examples are graph coloring \cite{jia2004much} and quasi-group completion \cite{gomes1997heavy}.
In this work, the GP distribution is used in lieu of the Pareto distribution. It is a stronger type of distribution which includes 
the Pareto distribution and is suitable to fit the tail behavior due to the Pickands–Balkema–de Haan theorem \cite{balkema1974residual}.

Another often considered distribution is the exponential distribution. However, both the Weibull and the GP are generalizations of the exponential distribution.

We are interested in the calculation of optimal restart times. Lorenz \cite{lorenz_runtime_2018} describes a method to find
the optimal restart times for all three distribution types. However, he found that for the Weibull distribution and the GP distribution 
the optimal approach is either instantly restarting or not restarting at all depending on the parameters. In practice instantly restarting results in much worse
performances. Therefore, the Weibull and the GP distribution are employed with a location parameter that shifts the whole distribution. This already resolves the paradox of instantly restarting, since the resulting (theoretically) optimal restart times are greater
than the location parameter.
%\begin{figure}[htbp]
%	\input{images/logn_fit}
%	\caption{The empirical distribution function and the lognormal fit of a typical instance.}
%	\label{fig:compare_logn}
%	% logn: k3-n2500-m10546-r4.218-s2394072653
%\end{figure}
The used RTDs are obtained from the empirical distribution with maximum likelihood estimations (fits). 

\begin{figure}[htbp]
	\floatbox[{\capbeside\thisfloatsetup{capbesideposition={left,top},capbesidewidth=3.5cm}}]{figure}[\FBwidth]
	{\caption{The empirical distribution function and the Weibull fit. Here, the Weibull distribution does not capture
			the runtime behavior well.}\label{fig:weib_bad}}
	{\input{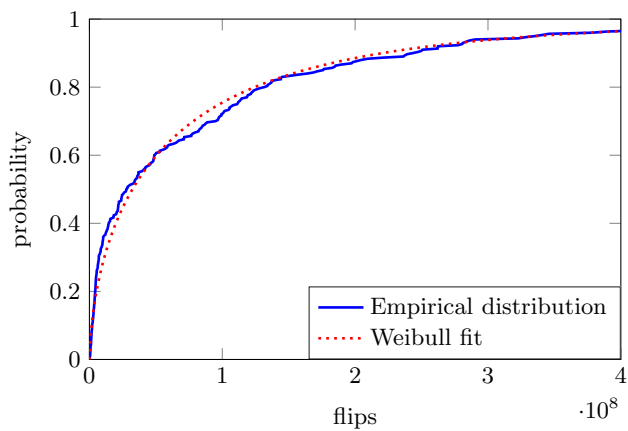}}
%	\caption{The empirical distribution function and the Weibull fit. Here, the Weibull distribution does not capture
%		the runtime behavior well.}
%	\label{fig:weib_bad}
	% weib: k3-n1500-m6361-r4.241-s2879353699
\end{figure}

Figure \ref{fig:weib_bad} shows the empirical RTD and a Weibull fit. Here, the Weibull distribution is the best fitting distribution, but
there are some substantial differences between the observed and the fitted runtime behavior. The empirical RTD shows a steeper ascend in the beginning but also shows some irregularities at about $10^8$ flips. For the sake of demonstration, we picked an example where the fit does not describe the empirical data well. Most other instances are described better by their respective best fit.
 
A behavior as especially in Figure \ref{fig:weib_bad} can lead to predictions of the (theoretical) optimal restart time which are far from the actual optimal restart time. These results are discussed in more detail in the next section. An advantage of using fitted distributions is that the complete runtime behavior can be expressed in just two parameters for the lognormal distribution or three parameters for the Weibull and GP distribution with location parameter.

%\begin{figure}[htbp]
%	\input{images/gp_fit}
%	\caption{The pdf of the lognormal distribution with a constant scale of 0 and variable shapes on the left, and constant shape of 1 and variable scales on the right.}
%	\label{fig:compare_gp}
%	% gp: k3-n2000-m8482-r4.241-s872520864
%\end{figure}

%  Barrero et al. [4] studied generation based models
%without selective pressure, they argue that the generations-to-success can be modeled
%by a Weibull distribution.

%Sometimes a probability distribution describes the runtime behavior of an algorithm well only if 
%the distribution is shifted. If the cumulative density function (cdf) $F(x)$ is substituted by $F(x-x_0)$, then the distribution is shifted by $x_0$.
%In the following, $x_0$ is called the location parameter. In this work, the Weibull distribution and the GP distribution use a location parameter while the lognormal distribution is employed without it. 
%
%All presented distribution types can describe heavy-tail phenomena which are tails that decay slower than an exponential function. Heavy-tails often indicate the beneficial application of restarts.

%Another type of distribution should also be considered in this context: The exponential distribution. 
%It has been considered for several algorithms including probsat on 5-SAT random instances \cite{arbelaez2016learning}.
%However, the exponential distribution can be modeled by both the Weibull distribution and the generalized Pareto distribution, therefore, 
%it is omitted in our analysis.  

\section{The Potential of Restarts in ProbSAT}
Stochastic local search (SLS) algorithms commonly employ restart strategies when the number of local search steps exceeds a cutoff value. However, finding a choice for the cutoff value is a hard task and strategies which work well for one instance might be a bad choice for other instances. 
However, if the runtime behavior always follows a particular distribution type and only the parameters of the distribution vary, then several conclusions about useful restart strategies can be drawn.
Thus, it can be useful to study the algorithm's runtime behavior for a specific problem class and model the behavior with RTDs.

Here, we study an SLS SAT-solver called probSAT \cite{balint_choosing_2012}. Information on the details of the probSAT algorithm can be found in Appendix \ref{sec:pSAT}. While probSAT is simple to implement it excels on random SAT formulas close to the phase transition. This can be seen in the parallel track of the SAT competition 2014 where a parallel version of probSAT \cite{balint2014probsat} won. The probSAT algorithm is also used in the SAT-solver dimetheus \cite{gableske2016sat} which is the best performing solver in the random track of the SAT competition in recent years. 

The original version of probSAT does not restart, while the parallel version of probSAT  uses one process with restart time $10^7$ flips and another process with restart time $10^8$ flips regardless of the size and the structure of the instance. The authors did not explain how
this restart time was obtained. Thus, in this section, we systematically analyze the runtime behavior and show that 
there is still much optimizing potential in the probSAT algorithm by using a carefully crafted restart strategy.
Furthermore, the analysis implies that the runtime behavior of probSAT can be well described by three distributions: The lognormal, the Weibull and the GP distribution. In fact, two distribution types suffice to calculate restart times which significantly improve the performance of probSAT.

Here, we study a version of probSAT which is known as the break-only-poly-version of probSAT\footnote{The original version can be found here: \url{https://github.com/adrianopolus/probSAT}.} with $c_b=2.3$ which is suggested by Balint and Schöning \cite{balint_choosing_2012} for random 3-SAT formulas. The used implementation is extended by the possibility to use Luby's strategy and to use a timeout after a fixed number of flips.

%We know that given the RTD we can calculate an optimal restart time with respect to the expected runtime. However, empirical data is almost never perfectly represented by such a distribution. Therefore calculating a restart time from a distribution fitted to the measured runtimes will be off of the optimal restart time.
%
%In this section, we analyze how much performance is lost by calculating the restart time from a distribution fitted to observed data compared to restarting at the optimal restart time.
%We use the break-only-poly-version of probSAT\footnote{The original version can be found here: \url{https://github.com/adrianopolus/probSAT}.} with $c_b=2.3$ as suggested by Balint and Schöning \cite{balint_choosing_2012}.
% As motivated above, we will consider the lognormal, Weibull and GP distribution. We compare the restart times based on their speedup to running the algorithm without any restarts.

\subsection{Instance specification}
\label{sec:instances}
For the experiments, random 3-SAT instances ranging from 1500 to 2500 variables are created using the generator kcnfgen \cite{gableske2016sat} by Oliver Gableske. We decided to use random instances since they are the typical use case of SLS-solvers.

The number of clauses was chosen such that the clause to variable ratio is close to the phase transition at about 4.27, but mostly below this ratio, since only satisfiable formulas are of interest.
 We stayed slightly below the phase transition because the instances in this range tend to be ``interesting'' as they are satisfiable but often hard to solve. Formulas with lower ratios, while almost always satisfiable, are mostly trivial to solve. Above the phase transition, the formulas tend to be unsatisfiable. 
 Most SLS solvers and especially probSAT are so-called incomplete solvers, i.e., if a formula is unsatisfiable, they do not terminate. 
 Thus, the RTDs can only be studied on satisfiable instances.
 The smallest used ratio was 4.204, the largest 4.272.

The complete instance set consists of 2400 formulas, 1632 of which are satisfiable, ensured by the SAT-solver dimetheus \cite{gableske2016sat}. Note that this can be seen as filtering since formulas on which dimetheus performs well are preferred. Only the 1632 formulas guaranteed to be satisfiable are used for the further steps. We decided to use dimetheus for this task since it is currently one of the best performing solvers on random instances.

\subsection{Empirical Distributions}
\label{sec:empirical}

For an instance $i$, the goal is to find a good approximation of its RTD $X_{i}$ on probSAT. For this, we sample the random variable $X_{i}$ 300 times to ensure stable results, measuring the number of flips until a satisfying assignment is found. This measure is chosen because it is stable and independent of hardware and scheduling. From this, we gain an empirical distribution function $\hat{F}_{300}(X_{i})$. The solving is done with the help of Sputnik by V\"{o}lkel et al. \cite{volkel2014sputnik}. 
The maximum likelihood fits on the empirical distribution yield an estimation of the parameters for a lognormal, a Weibull and a GP distribution.
The Kolmogorov-Smirnov test (KS-test) is applied to compare these new distributions with $X_i$.
%To compare these new distributions with $X_{i}$ we apply the Kolmogorov-Smirnov test (KS-test). 
Note, that passing the KS-test is not a proof of that distribution being the correct one. We only use it as a goodness-of-fit argument to see how well the actual distribution is described.

Only ten out of all 1632 instances could not be classified as any of the distributions according to the KS-test at a significance level of 0.05.
%Out of all 1632 instances, for only 10 none of the distributions passed the KS-test at a significance level of 0.05. 
%As a label for the later training we designated a "winning" distribution by the highest p-value. 
The distribution with the highest p-value is designated as the ``winning'' distribution.
Table \ref{tab:ksresults} displays how often the distributions won or passed the KS-test.
%How often the distributions won or passed the KS-test is displayed in Table \ref{tab:ksresults}.

\begin{table}[ht]
\centering
	\floatbox[{\capbeside\thisfloatsetup{capbesideposition={left,top},capbesidewidth=6.3cm}}]{table}[\FBwidth]
	{\caption{The number of instances where the fit of the respective distribution passed the KS-test at a significance level of 0.05. A distribution ``won'' the test, if it passed with the highest p-value.}\label{tab:ksresults}}
%$\begin{array}{|c|c|c|c|c|}
{\setlength{\tabcolsep}{5pt}
\begin{tabular}{l|llll}
	%\hline
	& \mbox{W} & \mbox{L} & \mbox{GP} & \mbox{none}\\
	\hline
	\mbox{Passed} & 1529 & 1273 & 1382 & 10\\
	%\hline
	\mbox{Won} & 541 & 606 & 475 & 10\\
	%\hline
\end{tabular}
}
%$\begin{array}{l|cccc}
%%\hline
%& \mbox{Weibull} & \mbox{lognormal} & \mbox{GP} & \mbox{none}\\
%\hline
%\mbox{Passed} & 1529 & 1273 & 1382 & 10\\
%%\hline
%\mbox{Won} & 541 & 606 & 475 & 10\\
%%\hline
%\end{array}$
%\caption{The number of instances where the maximum-likelihood fit of the respective distribution passed the KS-test at a significance level of 0.05. A distribution ``won'' the test, if it passed with the highest p-value.}
%\label{tab:ksresults}
\end{table}

%\vspace{-1em}

It is noticeable that the Weibull distribution (W) describes about 93.7\% of the observed RTDs well. This value increases to above 97\% at a significance level of 0.01.

Furthermore, 78.0\% of the instances can be described by a lognormal distribution (L). This supports the observations of Arbelaez et al. \cite{arbelaez_using_2013}, who report that 389 of their 500 instances (77.8\%) are described well by the lognormal distribution, with the same approach and at a significance level of 0.05.

On a side note, by applying the same procedure to the (shifted) exponential distribution, we observed that it covers 67.7\% of the instances in the shifted and only 52.3\% in the unshifted case.

\subsection{Calculating the Potential}

We approximate an optimal expected runtime using the following approach. Let $X$ be the set of observed runtimes of a fixed instance, and $X_{\leq x}$ the set of elements from $X$ that are smaller than $x$. Then
\begin{equation}
	\hat{opt} = min_{x\in X}\left(\frac{1-p}{p}\cdot x + \frac{\sum_{\{y \in X_{\leq x}\}}y}{|X_{\leq x}|}\right)
\end{equation}
with $p = \frac{|X_{\leq x}|}{|X|}$. This is the minimum expected runtime if restarted at an observed runtime given the measured data.
A standard tool to compare performances is given by the speedup. Let $Y_k$ be any fixed instance and $Y_k^{(i)}$ be the runtime of algorithm $i$ on instance $Y_k$. Then the speedup $s_{Y_k}^{(i,j)}$ is defined as:
\begin{equation}
	s_{Y_k}^{(i,j)}= \frac{E[Y_k^{(i)}]}{E[Y_k^{(j)}]}.
\end{equation}
 The geometric mean $GM = \left(\prod_{k=1}^{m} s_{Y_k}^{(i,j)}\right)^{\frac{1}{m}}$
%\begin{equation}
%	GM = \left(\prod_{k=1}^{l} s_{Y_k}^{(i,j)}\right)^{\frac{1}{m}}
%\end{equation}
is suitable to argue about the average speedup on $m$ instances with speedups $s_{Y_1}^{(i,j)}, \dots, s_{Y_m}^{(i,j)}$.
With this, the average speedup of using the optimal restart time compared to not restarting is 1.393 which serves as a baseline for further comparisons. We interpret this value as the maximum speedup reachable with a fixed-cutoff restart strategy. This value in itself is already a strong result: If one knew the optimal restart times, one could improve the performance of probSAT by about 39\%.

The speedup for the different distributions is calculated in a similar way. However, the considered restart time is now calculated from the parameters of the distribution as proposed by Lorenz in \cite{lorenz_runtime_2018}. We calculate the speedup for each subset of the set of distribution types \{lognormal (L), Weibull (W), GP\}. Within each subset, we distinguish between two strategies. 
The first uses the distribution that performs best under the KS-test (highest p-value). The other chooses the distribution with the highest speedup. 
For both cases, this is decided for each instance individually.
The results are displayed in Table \ref{tab:speedups}.

\begin{table}[ht]
\centering
	\floatbox[{\capbeside\thisfloatsetup{capbesideposition={left,top},capbesidewidth=5cm}}]{table}[\FBwidth]
	{\caption{The calculated speedups for different sets of probability distribution types. In sets with several distributions, the one was picked to calculate the speedup that performed better in the KS-test in the middle column, and that led to the higher speedup in the right column.}\label{tab:speedups}}
{$\begin{array}{l|ll}
	& \mbox{KS-test} & \mbox{speedup} \\
	\hline
	\emptyset (\mbox{base line}) & 1.393 & 1.393 \\
	\{L\} & 1.154 & 1.154 \\
	\{W\} & 1.174 & 1.174 \\
	\{GP\} & 1.157 & 1.157 \\
	\{L,W\} & 1.253 & 1.284\\
	\{L,GP\} & 1.162 & 1.184\\
	\{W,GP\} & 1.187 & 1.285\\
	\{L,W,GP\} & 1.200 & 1.300
\end{array}$}
%\caption{The calculated speedups for different sets of probability distribution types. In sets with several distributions, the one was picked to calculate the speedup that performed better in the KS-test in the middle column, and that led to the higher speedup in the right column.}
%\label{tab:speedups}
\end{table}

Clearly a lot of potential is already lost just by the representation of the data as fitted probability distributions. Even in the best case,
%of using the best speedup while all three distributions are considered, 
the speedup is lower by about 0.1 compared to the baseline. Additionally, we can see, that a single distribution type does not suffice to express the observed RTDs. Overall, the combination of Weibull and lognormal seems to be most stable, performing far better than any other combination in the KS-Test column, and being close to the best combination in the speedup column. 
%Lastly, while the GP distribution performed on comparable levels than the other distributions, pairing it up with the others performed worse than any compination not inluding the GP.

A reason for the observed discrepancy between the baseline and all other speedups could be that the 
fitted distributions are too smooth as can be seen in the figure discussed in section \ref{sec:bla}. Describing the runtime behavior with few parameters does not suffice 
to capture all necessary details. Furthermore, the used sampling method does not provide
any specifics on very short runs and the probabilities associated with them. As a side note, restarting at $10^7$ or $10^8$ flips as proposed in \cite{balint2014probsat} yields worse results than any of the distributions. For some instances the chosen restart time is too low by several magnitudes which deteriorates the performance.

\section{From Theory to Practice}
The previous sections argued about the speedup while estimating it from data obtained by running probSAT without any restarts. While this gives an idea of what is possible, it does not explain how we can use this information in practice. 
It is possible to model the estimation of the optimal restart time as a regression problem. Nevertheless, this approach is limited: 
Shylo et al. \cite{shylo_restart_2011} found that optimal restart times are generally increasing with an increasing number of processors.
Since the regressor is trained with data obtained from a fixed number of processors, its estimations are only accurate for the same number of processors.
Estimating the RTDs is more flexible: 
The estimates can be easily used to find the optimal restart time on an arbitrary number of processors. 
Other scenarios can also benefit from the fits, like predicting the probability of completion within a deadline.
We choose to predict the RTDs since it is a flexible model.
However, obtaining the parameters of the fitted distribution is a non-trivial task. Eggensperger et al. \cite{eggensperger_predicting_2017} propose a machine learning pipeline for this task, which we adapt to fit our setting.

There are two steps in the estimation of the parameters:
First, a random forest estimates the type of the distribution. Second, based on the type, a neural network trained specifically for that distribution type predicts the distribution's parameters.

We experimented with several combinations of distribution types and found that the combination of lognormal and Weibull leads to the best results. While the previously observed potential speedup of the GP distribution is comparable to the other distributions, its network performed significantly worse. For this reason, the upcoming experiments omit the GP distribution.

We chose the final model of each component by comparing its performance on a test set. However, we did not measure its performance based on the respective loss function, but on the potential speedup, calculated in the same way as described in the previous section. 
The potential speedup is not suitable as a loss function for technical reasons.
A lot of extra information is required to calculate the speedups, leading to very high runtimes in the learning process. 
The networks predict the distribution instead of the restart times since this work focuses on how well probSAT performs with restart times calculated from
fitted RTDs.

The remainder of this section describes the applied features and details of the machine learning components. The quality of the components is measured on a test set. The test set consists of a sample of 162 instances where instances with exceptionally high or low shape parameters are oversampled on purpose. These were sampled from the 1632 instances mentioned in section \ref{sec:instances}. During training, only the remaining 1470 instances were used.

\subsection{Feature Extraction}
The SATzilla feature extractor by Xu et al. \cite{xu2012satzilla2012} creates the features of the instances (as motivated by Arbelaez et al. \cite{arbelaez2016learning}).
It is called with the parameters -base, -ls and -lobjois, leading to a total of 81 features. A description of the features is found in \cite{xu2012features}. After normalizing the features to have minimum 0 and maximum 1, we selected the features with a variance larger than 0.05 and an additional four handpicked features. A total of 34 features remained. The used features are provided in the appendix.

\subsection{Random Forest}

The distribution type of an instance is estimated by a random forest. The label is the distribution that performed better under the KS-test (higher p-value).
The random forest does not use normalized features.
Normalization was tested as well, but no significant difference in
the result was observed.
The random forest was evaluated by ten iterations of a 10-fold cross-validation. The average potential speedup measured during the cross-validation is 1.218, further details of the evaluation of the random forest can be found in the appendix. We implemented the random forest with Python's scikit-learn \cite{scikit-learn} library.

\subsection{Neural Networks}
% weib-cross val: 1.1366
% log-cross val: 1.1270
Neural networks (NNs) predict the shape, scale and location parameter of the distributions, whereas the location parameter is only predicted for the Weibull distribution.
Both distribution types use a separate NN. 
The experiments indicate that the inclusion of the location parameter into the network with the other parameters significantly worsens the overall performance.
Therefore we trained a separate network to predict the location parameter.
The networks are implemented in Python with the Keras framework \cite{chollet2015keras} and tensorflow backend \cite{tensorflow2015-whitepaper}. The specifications of the networks can be found in the appendix.

A standard loss function is the root-mean-square error (RMSE) of a prediction wrt. the ground truth. It was applied in the location network.
Nevertheless, in the other NNs, two parameters have to be trained. It is possible to measure the 
sum of the two RMSE values for shape and scale, but the shape and the scale parameters are clearly related, thus
the relationship between both parameters would be lost. Hence, it is advisable to use a loss function which captures the
relationship between both parameters.
Eggensperger et al. \cite{eggensperger_predicting_2017} use the negative log-likelihood function for measuring the quality of a fit. We also experimented with this function but 
achieved better results by using a different loss function based on the KS-Test. 
The loss function is presented in equation \ref{eq:loss}. Here, $\sigma$ is the shape and $\mu$ the scale parameter.
The labels are the value of the maximum-likelihood estimation for $\sigma$, one observed runtime $x$ and the associated probability of the empirical distribution:
\begin{equation}
\label{eq:loss}
	L(\hat{\sigma}, \hat{\mu}) = \Bigl| F(x\mid \hat{\sigma}, \hat{\mu})- \textnormal{empirical} \Bigl| + \Bigl| \sigma - \hat{\sigma} \Bigl| ,
\end{equation}
where $F$ is the corresponding cumulative distribution function.
Here, $\hat{\sigma}$ and $\hat{\mu}$ are the predictions by the NN. 
This means that the network tends to predictions which minimize the absolute difference between the cumulative distribution function and the empirical distribution.
It is also possible to define the loss function without the latter part $\vert \sigma - \hat{\sigma} \vert$, but in our experiments, this term improved the predictions of the shape parameters significantly. 
%Only the scale and the shape parameter are part of the NN. Including the location parameter into the network significantly worsened the predictions of both shape and scale, hence it is separately estimated by a linear regression model implemented with scikit-learn \cite{scikit-learn}.

Lorenz \cite{lorenz_runtime_2018} found that for all distributions the usefulness of restarts only depends on the shape parameters.
Thus, it is especially important to predict the shape parameters of extreme cases well for the intended use case.  

The performance of the NNs is checked and verified by ten iterations of a 10-fold cross-validation. The measured average potential speedups are 1.137 for the Weibull and 1.127 for the lognormal distribution. The average RMSE of the location network is 0.714. 

A remarkable behavior during the evaluation of the Weibull network for shape and scale is linked to the location parameter.
Estimating parameters that lead to a restart time that is below the location parameter drastically reduce the calculated speedup.
A reason for this is that below the location parameter there are no observed runtime available. Thus, the density of the RTD is unknown.
To be able to calculate any speedup,  we approximate the behavior by an exponential function, so the density tends towards zero fast when reaching values lower than the location parameter. During the above evaluations, instances for which the network calculated a restart time below the location parameter are omitted. However, this occurred only five times out of 162 instances.
Evaluating the whole pipeline on the test set lead to a remaining potential speedup of 1.157.

%%%%%%  CROSS- VAL: TEST SET:
%weib scale log rmse: 1.5752904812
%weib shape rmse: 0.0915338663654
%par scale log rmse: 2.36985485919
%par shape rmse: 0.153760195945
%logn scale rmse: 0.657745805787
%logn shape rmse: 0.105447217458

%%%%% CROSS-VAL: VALIDATION SET:
%weib scale log rmse: 1.31397858643
%weib shape rmse: 0.115072001414
%par scale log rmse: 2.91756233351
%par shape rmse: 0.214910456603
%logn scale rmse: 0.848368267994
%logn shape rmse: 0.144268850796

%%%% CROSS-VAL: KS-METRIC
%logn_test.txt :  3.8512046923
%logn_val.txt :  4.75961406743
%par_test.txt :  5.92163692402
%par_val.txt :  6.93821230728
%weib_test.txt :  4.30730930484
%weib_val.txt :  4.39661559255

\section{Experimental Results}
%  all
% t-test: new vs no: significant: 0.0044
% Wilcoxin Rank test: 0.0174
% Luby: 1.644187259413985e-50
% speedup: 

%% just restarts:
% t-test: 5.227904480000114e-04
% Wilcoxin Rank test: 9.8704e-04
% speedup: 1.107
We compare three different models: A fixed-cutoff strategy where the restart times are calculated by using the parameters from the NN. Note, that this method might lead to no restarts. In the following evaluation, this approach is denoted by ``static restarts''.
Secondly, Luby restarts are  considered. The Luby strategy theoretically optimal and thus a good baseline. 
When using Luby restarts, the $i$-th term $t_i$ can be calculated by the following formula.
\begin{equation}
	t_i =
	\left\{
		\begin{array}{ll}
			2^{k-1}  & \mbox{if } i=2^k-1 \\
			t_{i-2^{k-1}+1} & \mbox{else}.
		\end{array}
	\right.
\end{equation}
Usually, an initialization  term $a$ is multiplied on $t_i$ to obtain the restart time $T_i$.
We use Luby restarts initialized with $20n$, where $n$ is the number of variables, this restart time is suggested by Balint and Schöning \cite{balint_choosing_2012}. Finally, not restarting the algorithm is considered.

Each of the approaches described above is tested on 100 new satisfiable instances generated with the
same settings as defined in section \ref{sec:instances}. The runtimes for each instance are sampled 100 times, and the average runtimes of each strategy are compared. For this experiment, a timeout of $10^{11}$ flips was used for the static restart strategy. Only one instance was affected by the cutoff. For this instance the static restart strategy found a solution in 72 out of 100 runs, the Luby strategy found a solution in 47 cases and the not restarting approach in 28 cases. The following analysis
only considers the remaining 99 instances. Again, we measure the runtime by the number of variable flips until a satisfying assignment is found.

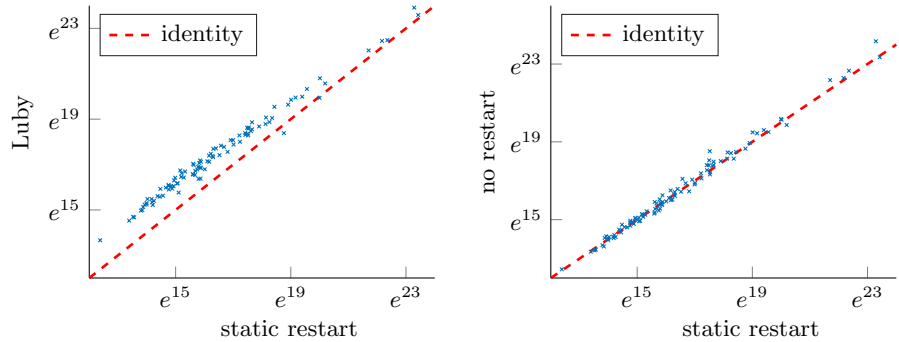
\begin{figure}[htbp]
	\begin{minipage}{0.49 \linewidth}
\definecolor{mycolor1}{rgb}{0.00000,0.44700,0.74100}%
\begin{tikzpicture}

\begin{axis}[%
width=4.521in*0.4,
height=3.566in*0.4,
at={(0.758in,0.481in)},
scale only axis,
xmin=12,
xmax=24,
ymin=12,
ymax=24,
xtick={15,19,23},
xticklabels={$e^{15}$,$e^{19}$, $e^{23}$},
ytick={15,19,23},
yticklabels={$e^{15}$,$e^{19}$, $e^{23}$},
y label style={at={(axis description cs:0.13,.57)},anchor=south},
ylabel=Luby,
x label style={at={(axis description cs:0.58,-.1)},anchor=south},
xlabel=static restart,
axis background/.style={fill=white},
axis x line*=bottom,
axis y line*=left,
legend style={at={(0.03,0.97)}, anchor=north west, legend cell align=left, align=left, draw=white!15!black}
]
\addplot[only marks, mark=x, mark options={}, mark size=1.000pt, draw=mycolor1, forget plot] table[row sep=crcr]{%
x	y\\
15.0607359616869	16.1801264647996\\
16.3112238880052	17.6940977847449\\
15.2077193592501	16.6126751432965\\
14.9266944188966	16.0467184563737\\
15.1930595874028	16.4576716694328\\
14.9432349018481	16.2975803643811\\
19.5604719619599	20.3289149118111\\
16.8053173514854	17.559105616813\\
17.519616228661	18.2825823070395\\
14.1854816678161	15.337030786837\\
13.9907657064001	15.3030113025202\\
18.1229525756381	18.7840202532173\\
14.470850990451	15.9279919996629\\
23.4252039246038	23.5891701137498\\
16.9942624196317	18.0733859478342\\
13.3877965126824	14.5271763113831\\
15.1038837323159	15.7643471849884\\
16.1670995798975	17.1247791084805\\
17.6533707334571	18.8603225964694\\
16.0113106912616	16.7900642032468\\
18.2579588064381	18.8681499362454\\
15.750150098798	16.7783186074472\\
17.6151038405536	18.4702472749568\\
16.6828489819472	17.7772712322518\\
23.2873615738876	23.9148887492358\\
18.3527096133461	19.0468305012044\\
17.5359278900936	18.6292377898075\\
17.4174544030363	18.3878861155145\\
16.1605560270782	17.3934029936879\\
15.8424718511051	16.8606026257986\\
14.559826742438	15.6271689766468\\
22.3553729342425	22.4853632382281\\
16.4007921927795	17.7124216380682\\
13.9893535648872	15.241465999204\\
14.2057101859397	15.2182878707809\\
13.8726350284504	15.1317417732722\\
17.2147934251125	18.2839259945765\\
14.6555596091006	15.9834835339901\\
15.3281070109836	16.6915419435833\\
17.6556013926804	18.5499319515774\\
14.9855034013337	16.4197533692766\\
16.5238737634088	17.4119570699279\\
15.7970172440147	16.3691457727578\\
16.573052706135	17.8316016669254\\
17.9005258305455	18.6774639546568\\
21.7014821924752	22.0285302843318\\
14.7595461955367	16.0584253238453\\
19.998506945499	19.936463548311\\
16.9109610710814	18.071784481473\\
18.9081513321738	19.6395932827658\\
18.4305752813679	19.5458513936356\\
12.3808870943697	13.6630222705967\\
14.7547456629532	16.1014576193629\\
20.0104629718737	20.8010913491531\\
14.8825578527334	15.9416456173857\\
17.4833195717566	18.6169070652989\\
15.1610047941164	16.7392985333176\\
15.8790102953948	16.8548728272703\\
14.7650109032467	15.9012141693949\\
20.1922391472391	20.5734022171513\\
16.2203394006433	17.4769747320905\\
15.8781423974261	17.1250841107072\\
15.6828495419293	16.7211024787493\\
17.012059044574	17.8809272888591\\
18.7594795491556	18.3828701178948\\
13.9247396908633	15.2303589788769\\
22.1682683499822	22.4465349766903\\
15.282982859829	16.4262081344912\\
15.6104564036321	17.0345117257767\\
19.3977729294802	19.9897365378358\\
17.525300184888	18.5868600157013\\
17.4241600280154	18.3087299853801\\
15.8634491801557	16.3804507315375\\
15.8205989549381	17.0616842002936\\
13.7951167405496	14.9707065557316\\
14.3710587702331	15.6072109921165\\
14.8459489539764	16.0490108634615\\
18.1392240406829	19.0735147092919\\
15.6715252300741	16.5418217646874\\
15.594115019928	16.5814513943965\\
19.0095346442117	19.8556787212584\\
14.2479066107518	15.4036476317839\\
19.1647848525525	19.9509751394914\\
16.2811830006747	17.3777228625398\\
15.6254686121642	16.9801698506826\\
17.2335094720605	18.002043129266\\
15.6105526734089	16.5614298181212\\
16.3026009599622	17.3237501684117\\
14.1652029857778	15.4820850168881\\
13.5200280074023	14.6938611803621\\
15.0638482206502	16.1588919289879\\
13.8419363409175	14.993342252409\\
14.4449774250211	15.5803634826205\\
15.8490211364361	17.1808426009827\\
13.5612007064507	14.6729811544566\\
15.7798156282761	16.9011569062267\\
16.1038250069677	17.1428632556811\\
17.5015497967325	18.364358952893\\
14.0219626344446	15.4814678395372\\
};
\addplot [color=red, dashed, line width=1pt]
  table[row sep=crcr]{%
12	12\\
24	24\\
};
\addlegendentry{identity}
\end{axis}
\end{tikzpicture}%
	\end{minipage}
	\hfill
	\begin{minipage}{0.49 \linewidth}
\definecolor{mycolor1}{rgb}{0.00000,0.44700,0.74100}%
\begin{tikzpicture}

\begin{axis}[%
width=4.521in*0.4,
height=3.566in*0.4,
at={(0.758in,0.481in)},
scale only axis,
xmin=12,
xmax=24,
ymin=12,
ymax=26,
xtick={15,19,23},
xticklabels={$e^{15}$,$e^{19}$, $e^{23}$},
ytick={15,19,23},
yticklabels={$e^{15}$,$e^{19}$, $e^{23}$},
y label style={at={(axis description cs:0.13,.5)},anchor=south},
ylabel=no restart,
x label style={at={(axis description cs:0.58,-.1)},anchor=south},
xlabel=static restart,
axis background/.style={fill=white},
axis x line*=bottom,
axis y line*=left,
legend style={at={(0.03,0.97)}, anchor=north west, legend cell align=left, align=left, draw=white!15!black}
]
\addplot[only marks, mark=x, mark options={}, mark size=1.000pt, draw=mycolor1, forget plot] table[row sep=crcr]{%
x	y\\
15.0607359616869	15.1056267166572\\
16.3112238880052	16.2167272623331\\
15.2077193592501	15.1464615103608\\
14.9266944188966	15.0865261968905\\
15.1930595874028	15.2558582442053\\
14.9432349018481	15.0604967887262\\
19.5604719619599	19.5007901021424\\
16.8053173514854	16.4556957534425\\
17.519616228661	18.5207360965049\\
14.1854816678161	14.0520346768695\\
13.9907657064001	14.0192454131812\\
18.1229525756381	18.1359011912451\\
14.470850990451	14.7416940546783\\
23.4252039246038	23.3504188750641\\
16.9942624196317	16.808397594668\\
13.3877965126824	13.350848906912\\
15.1038837323159	14.9386916726997\\
16.1670995798975	16.0089194102276\\
17.6533707334571	17.9793489447532\\
16.0113106912616	15.9412294486219\\
18.2579588064381	18.4212008058967\\
15.750150098798	15.4687477110796\\
17.6151038405536	17.7588801859479\\
16.6828489819472	16.8679381394468\\
23.2873615738876	24.1705334588894\\
18.3527096133461	18.1344577316212\\
17.5359278900936	17.4356165372363\\
17.4174544030363	17.8008269454005\\
16.1605560270782	16.5013161439837\\
15.8424718511051	15.8229787136011\\
14.559826742438	14.6130553567451\\
22.3553729342425	22.6576676614715\\
16.4007921927795	16.5805245114426\\
13.9893535648872	13.949631751622\\
14.2057101859397	14.2104650963815\\
13.8726350284504	14.0597517721638\\
17.2147934251125	17.3810240527718\\
14.6555596091006	14.6015890228172\\
15.3281070109836	15.427363276406\\
17.6556013926804	17.8117854863383\\
14.9855034013337	15.0040154109571\\
16.5238737634088	16.4379997719346\\
15.7970172440147	15.5672442676762\\
16.573052706135	17.0997070708727\\
17.9005258305455	18.010790871296\\
21.7014821924752	22.1745067168123\\
14.7595461955367	14.6895966357181\\
19.998506945499	20.1781213733315\\
16.9109610710814	17.0560013088835\\
18.9081513321738	18.9397372713023\\
18.4305752813679	18.4846811238804\\
12.3808870943697	12.4480706100217\\
14.7547456629532	14.8537000221618\\
20.0104629718737	20.1353783553965\\
14.8825578527334	14.7859663320083\\
17.4833195717566	17.7995236883043\\
15.1610047941164	15.2823292223104\\
15.8790102953948	15.9073987045378\\
14.7650109032467	14.944200827993\\
20.1922391472391	19.8670463290629\\
16.2203394006433	16.0238841072323\\
15.8781423974261	16.2513556405064\\
15.6828495419293	15.5951992307068\\
17.012059044574	16.8918151321635\\
18.7594795491556	18.650313960085\\
13.9247396908633	14.1466172471749\\
22.1682683499822	22.2853329654235\\
15.282982859829	15.2548414190806\\
15.6104564036321	15.9093628455147\\
19.3977729294802	19.6033262758438\\
17.525300184888	17.3276072451285\\
17.4241600280154	17.5430816208574\\
15.8634491801557	15.7494662752542\\
15.8205989549381	15.8685359425414\\
13.7951167405496	13.68743322488\\
14.3710587702331	14.4901709357511\\
14.8459489539764	14.8952416059292\\
18.1392240406829	18.4565126715563\\
15.6715252300741	15.8327903264688\\
15.594115019928	15.3124812574375\\
19.0095346442117	19.4884261917209\\
14.2479066107518	14.1012310978683\\
19.1647848525525	19.4388050189185\\
16.2811830006747	16.2991601855308\\
15.6254686121642	15.5486654369668\\
17.2335094720605	17.1199484231743\\
15.6105526734089	15.3697635777661\\
16.3026009599622	16.4217330645298\\
14.1652029857778	14.1204490040522\\
13.5200280074023	13.4204658929615\\
15.0638482206502	14.9200451683473\\
13.8419363409175	13.6213135462544\\
14.4449774250211	14.3445994129994\\
15.8490211364361	15.8146674510386\\
13.5612007064507	13.4475341639646\\
15.7798156282761	16.0423220719898\\
16.1038250069677	16.1082446083759\\
17.5015497967325	18.0484849609618\\
14.0219626344446	14.1115077636759\\
};
\addplot [color=red, dashed, line width=1pt]
  table[row sep=crcr]{%
12	12\\
24	24\\
};
\addlegendentry{identity}
\end{axis}
\end{tikzpicture}%
	\end{minipage}
	\caption{On the left, the static restart strategy is compared with the Luby strategy. On the right, the static restart strategy is compared with not restarting. The static restart strategy is more efficient if the dot lies above the identity line; otherwise, its competitor is more efficient. The axes are log-scaled average runtimes measured in variable flips.}
	\label{fig:compare_mean}
\end{figure}

The comparison of the logarithmically scaled runtimes between not restarting, static restarts and Luby restarts is illustrated in Figure \ref{fig:compare_mean}. 
The static restart strategy clearly outperforms the Luby strategy on all but two instances. The comparison
with not restarting is harder to interpret. It can be seen that for easy instances there 
is barely any difference in the performance of the strategies. However, for intermediate and hard instances 
the static restart strategy outperforms not restarting in most cases. 
The parameters of probSAT were tuned without restarting with a short timeout. Therefore, it is not
surprising that easy instances do not profit from restarting. Hard instances, on the other hand,
can potentially show a heavy-tailed behavior, i.e., probSAT's parameters are not chosen optimally for the
case without restarts. 

The comparison between the static and the Luby strategy yields an average speedup of $2.725$.
The comparison between the static restart strategy and not restarting yields an overall average speedup of $1.063$. 
This speedup includes instances where the static strategy predicted not restarting. If only instances are considered where the static strategy restarted,
then an average speedup of $1.108$ is achieved on the remaining $69$ instances.

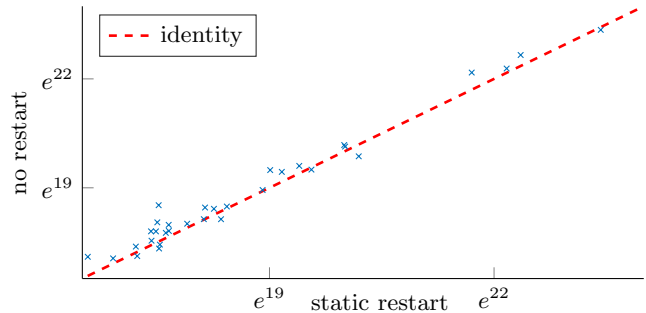
\begin{figure}[htbp]
	\floatbox[{\capbeside\thisfloatsetup{capbesideposition={left,top},capbesidewidth=3.3cm}}]{figure}[\FBwidth]
	{\caption{The 33 instances with the longest runs for which restarts are predicted. The axes are log-scaled average runtimes measured in variable flips.}\label{fig:compare_hard}}
	{% This file was created by matlab2tikz.
\definecolor{mycolor1}{rgb}{0.00000,0.44700,0.74100}%
\begin{tikzpicture}

\begin{axis}[%
width=4.521in*0.65,
height=3.566in*0.4,
at={(0.758in,0.481in)},
scale only axis,
xmin=16.5,
xmax=24,
ymin=16.5,
ymax=24,
xtick={19,22},
xticklabels={ $e^{19}$, $e^{22}$},
ytick={19,22},
yticklabels={$e^{19}$, $e^{22}$},
y label style={at={(axis description cs:0.08,.5)},anchor=south},
ylabel=no restart,
x label style={at={(axis description cs:0.53,.0)},anchor=south},
xlabel=static restart,
axis background/.style={fill=white},
axis x line*=bottom,
axis y line*=left,
legend style={at={(0.03,0.97)}, anchor=north west, legend cell align=left, align=left, draw=white!15!black}
]
\addplot[only marks, mark=x, mark options={}, mark size=1.500pt, draw=mycolor1, forget plot] table[row sep=crcr]{%
x	y\\
19.5604719619599	19.5007901021424\\
17.519616228661	18.5207360965049\\
18.1229525756381	18.1359011912451\\
23.4252039246038	23.3504188750641\\
17.6533707334571	17.9793489447532\\
18.2579588064381	18.4212008058967\\
17.6151038405536	17.7588801859479\\
23.2873615738876	24.1705334588894\\
18.3527096133461	18.1344577316212\\
17.5359278900936	17.4356165372363\\
17.4174544030363	17.8008269454005\\
22.3553729342425	22.6576676614715\\
17.2147934251125	17.3810240527718\\
17.6556013926804	17.8117854863383\\
16.573052706135	17.0997070708727\\
17.9005258305455	18.010790871296\\
21.7014821924752	22.1745067168123\\
19.998506945499	20.1781213733315\\
16.9109610710814	17.0560013088835\\
18.9081513321738	18.9397372713023\\
18.4305752813679	18.4846811238804\\
20.0104629718737	20.1353783553965\\
17.4833195717566	17.7995236883043\\
20.1922391472391	19.8670463290629\\
22.1682683499822	22.2853329654235\\
19.3977729294802	19.6033262758438\\
17.525300184888	17.3276072451285\\
17.4241600280154	17.5430816208574\\
18.1392240406829	18.4565126715563\\
19.0095346442117	19.4884261917209\\
19.1647848525525	19.4388050189185\\
17.2335094720605	17.1199484231743\\
17.5015497967325	18.0484849609618\\
};
\addplot [color=red, dashed, line width=1pt]
  table[row sep=crcr]{%
16	16\\
24	24\\
};
\addlegendentry{identity}
\end{axis}
\end{tikzpicture}%
}
%	\caption{The 33 instances with the longest runs for which restarts are predicted.}
%	\label{fig:compare_hard}
\end{figure}

Figure \ref{fig:compare_hard} shows that the speedup scales well with the expected runtime: 
When considering the $33$ instances with the longest runtimes, a speedup of $1.216$ is obtained.
Finally, the speedups are tested with the t-test and a modified Wilcoxon signed-rank test (\cite[Chapter 7.12]{hollander2013nonparametric}).
%When all instances are considered, the static strategy performs statistically significantly better than both the Luby strategy (t-test $p=1.644\mathrm{e}{-50}$, Wilcoxon $p$ below precision) and not restarting (t-test $p=0.0044$, Wilcoxon $p=0.0174$).
When all instances are considered, the static strategy performs statistically significantly better than both the Luby strategy (t-test
$p<2\mathrm{e}{-50}$, Wilcoxon $p<1\mathrm{e}{-50}$) and not restarting (t-test $p=0.0044$, Wilcoxon $p=0.0174$).
%When only instances are considered for which restarts are applied, then the static strategy is still significantly better than Luby's strategy (t-test $p=2.935\mathrm{e}{-35}$, Wilcoxon $p=2.720\mathrm{e}{-13}$) and not restarting (t-test $p=5.227\mathrm{e}{-4}$,Wilcoxon $p=9.870\mathrm{e}{-4}$).
When only instances are considered for which restarts are applied, then the static strategy is still significantly better than Luby's strategy (t-test
$p<3\mathrm{e}{-35}$, Wilcoxon $p<3\mathrm{e}{-13}$) and not restarting (t-test $p<6\mathrm{e}{-4}$,Wilcoxon $p<1\mathrm{e}{-3}$). 
%However, we observed an average speedup of $1.242$ for instances with an average runtime of more than $exp(17)$ flips (with restarts), these are $33$ of the $100$ instances. 
%The rest of the instances show an average speedup of just $1.016$. 
%The result becomes more obvious if even extremer instances are used for the calculation of the average speedup. We observed a speedup of $1.326$ on the ten hardest instances and a speedup of $1.674$ on the five hardest instances. 
%Furthermore, the speedup is tested with the t-test. The speedups compared to the Luby strategy ($p=9.93\mathrm{e}{-53}$) and compared to not restarting ($p=9.87\mathrm{e}{-4}$) are both statistically significant.
We conclude that our system combines the advantages of not restarting on easy instances while
also significantly improving the performance on hard problems.

%\vspace{-2em}

\section{Conclusion \& Outlook}
%\vspace{-1em}
This work analyzes the potential speedup of restarts for the probSAT algorithm on random 3-SAT instances close to the phase transition. A major result is that if the best restart time is chosen according to the data, then an average speedup factor of $1.393$ will be obtained. We proceed to approximate the runtime behavior with continuous probability distributions.  Three distribution types, the Weibull, the generalized Pareto and the lognormal distribution, are identified which describe the runtime behavior well. The (shifted) Weibull distribution performs best among these three as it describes the runtime distributions of $93.7\%$ instances well. 
The representations are used to calculate the (theoretically) best restart times; these predictions are used to measure the potential speedup of probSAT. An average speedup factor of up to $1.300$ can be achieved. We find it surprising how much potential probSAT has with respect to restarts since
the used probSAT implementation is optimized towards utilizing the algorithm without any restarts regarding the expected runtime \cite{balint_choosing_2012}.
Modifications that lead to the runtime distribution having a heavy-tail more often could increase the expected runtime without restarts, 
but the potential speedup with restarts could be increased further. %be beneficial for strategies with restarts. 
A handle for this modification in probSAT is, for example, the function that chooses the next variable to flip.

Since the generalized Pareto distribution worsened the results it is omitted in the following.
The observations and the approximated runtime distributions are employed 
to train a random forest and neural networks. 
The random forest is used to distinguish between the distributions while the neural networks are used to predict the parameters of the distributions. 
The predictions are used to decide whether restarts are useful. If they are, then the optimal restart time
is calculated, and the fixed-cutoff strategy is used. Otherwise, probSAT is not restarted.
We compare this approach with the Luby strategy and not restarting. The observations show that our approach combines the good behavior of probSAT on easy and intermediate instances while considerably improving the performance on hard instances. The presented approach is statistically significantly better than both Luby's strategy
and not restarting. An average speedup of $1.063$ is obtained on all instances and an average speedup of $1.216$ on the $33$ of $99$ test instances with the longest runtime where restarts are applied.

While the presented procedure already demonstrates the high potential of the approach, there are still open ends for further research.
A different set of features might improve the quality of the predictions. Features which describe runs of probSAT could be suitable candidates.
Furthermore, the used restart strategy is very optimistic, assuming that the estimated restart time is, in fact, the optimal one. More sophisticated strategies that are aware of inaccuracies of the network could lead to further improvements, especially if the estimations are far off the actual values. 

Naturally, our approach is not limited to probSAT. 
Any Las Vegas algorithm can be analyzed and optimized with the same technique.
The results imply that well-performing algorithms can be further improved with our approach, especially on hard instances.

\section{Acknowledgements}
The authors acknowledge support by the state of Baden-W\"{u}rttemberg \linebreak{}through bwHPC. Furthermore, we thank Gunnar V\"{o}lkel for many insightful discussions in the field of statistics and support with Sputnik. 
%We also want to mention Oliver Gableske for providing the programs kcnfgen and dimetheus, and lastly Adrian Balint, who allowed the use and adaptation of his probSAT implementation. 

\bibliography{restart} 

\appendix

\section{Features}
All machine learning components use 34 features. The neural networks use them in a normalized form with minimum 0 and maximum 1. A description of the features can be found in \cite{xu2012features}.

The following are the features, grouped based on \cite{xu2012features}. The number in brackets indicates the number of corresponding features.

\begin{itemize}
\item Problem Size Features:
	\begin{itemize}
 	\item[(2)] Number of variables and clauses in the original formula
 	\item[(2)] Number of variables and clauses after simplification\\
 	\end{itemize}
 	
\item Variable-Clause Graph Features:
	\begin{itemize}
	\item[(3)] Variable node degree statistics (mean, min, max)
	\item[(3)] Clause node degree statistics (mean, min, max)\\
	\end{itemize}
	
\item Variable Graph Features:
	\begin{itemize}
	\item[(3)] Node degree statistics (mean, min, max)\\
	\end{itemize}
	
\item Clause Graph Features:
	\begin{itemize}
	\item[(2)] Node degree statistics (mean, max)\\
	\end{itemize}
	
\item Proximity to Horn Formula:
	\begin{itemize}
	\item[(2)] Number of occurences in a Horn clause for each variable (mean, min)\\
	\end{itemize}	
	
\item DPLL Probing Features:
	\begin{itemize}
	\item[(1)] Search space size estimate\\
	\end{itemize}
	
\item Local Search Probing Features, based on 2 seconds of runnich each of SAPS and GSAT:
	\begin{itemize}
	\item[(10)] Number of steps to the best local minimum in a run
	\item[(2)] Average improvement to best in a run
	\item[(3)] Fraction of improvement due to first local minimum\\
	\end{itemize}	
	
\item Timing Features:
	\begin{itemize}
	\item[(1)] CPU time require for feature computation
	\end{itemize}
\end{itemize}

The internal names of the features are:

\begin{verbatim}
nvarsOrig, nclausesOrig, nvars, nclauses, VCG-CLAUSE-mean, VCG-CLAUSE-min,
\end{verbatim}
\begin{verbatim}
VCG-CLAUSE-max, VCG-VAR-mean, VCG-VAR-min, VCG-VAR-max, HORNY-VAR-mean,
\end{verbatim}
\begin{verbatim}
HORNY-VAR-min, VG-mean, VG-min, VG-max, CG-mean, CG-max, CG-featuretime, 
\end{verbatim}
\begin{verbatim}
saps_BestSolution_Mean, saps_BestSolution_CoeffVariance, 
\end{verbatim}
\begin{verbatim}
saps_FirstLocalMinStep_Mean, saps_FirstLocalMinStep_CoeffVariance, 
\end{verbatim}
\begin{verbatim}
saps_FirstLocalMinStep_Median, saps_FirstLocalMinStep_Q.10, 
\end{verbatim}
\begin{verbatim}
saps_FirstLocalMinStep_Q.90, saps_BestAvgImprovement_Mean, 
\end{verbatim}
\begin{verbatim}
gsat_BestSolution_Mean, gsat_FirstLocalMinStep_Mean, 
\end{verbatim}
\begin{verbatim}
gsat_FirstLocalMinStep_CoeffVariance, gsat_FirstLocalMinStep_Median, 
\end{verbatim}
\begin{verbatim}
gsat_FirstLocalMinStep_Q.10, gsat_FirstLocalMinStep_Q.90, 
\end{verbatim}
\begin{verbatim}
gsat_BestAvgImprovement_Mean, lobjois-mean-depth-over-vars
\end{verbatim}

\section{Specifications and Validation of the Random Forest}

The parameters were set to \begin{verbatim}n_estimators = 50, criterion = "entropy",\end{verbatim}
 all other parameters were default.
 
 Validating the random forest by ten iterations of a 10-fold cross validation resulted in a average potential speedup of 1.218. Evaluating the final model on an independent set of 162 instances resulted in a speedup of 1.138. The evaluation of the forests on other metrics is displayed in Table \ref{tab:acc}. All values concern the Weibull distribution as the label.

\begin{table}[ht]
\centering
$\begin{array}{l|ll}
\mbox{metric} & \mbox{cross validation} & \mbox{final model}\\
\hline
\mbox{true positive rate} & 0.803 & 0.750 \\
\mbox{true negative rate} & 0.648 & 0.571 \\
\mbox{precision} & 0.764 & 0.697 \\
\mbox{negative prediction value} & 0.698 & 0.635\\
\mbox{fall-out} & 0.351 & 0.429 \\
\mbox{miss rate} & 0.197 & 0.25 \\
\mbox{accuracy} & 0.738 & 0.673 \\
\mbox{balanced accuracy} & 0.726 & 0.661 
\end{array}$
\caption{The evaluation of the random forest for different metrics. All values concern the Weibull distribution as the label.}
\label{tab:acc}
\end{table}

We were not able to get these values significantly higher. Especially an accuracy of below 75\% is clearly not a desirable result. However, we still used the random forest, since the potential speedup was still promising. An explanation for this is that often both lognormal and Weibull describe the distribution well and therefore produce good restart times. Hence, classfying the wrong distribution does not necessarily have a negative effect on the potential speedup.

\section{Specification of the Neural Networks}

All networks used Adam as the optimizer, with initial learning rate 0.0005 and clipnorm 0.5. The inputs were processed in batches of 16.

The parameters of the networks are displayed in Table \ref{tab:netParams}. Regarding the number of output neurons: While the relevant number of neurons was two, we had to use three in the lognormal and four in the Weibull network for technical reasons.

\begin{table}[ht]
\centering
$\begin{array}{l|llll}
\mbox{layer} & \mbox{parameter} & \mbox{location} & \mbox{Weibull} & \mbox{lognormal}\\
\hline
\mbox{input} & \mbox{neurons} & 34 & 34 & 34\\
& \mbox{gaussian noise} & 0.01 & 0.01 & 0.01\\
\hline
\mbox{hidden1} & \mbox{neurons} & 14 & 14 & 14\\
& \mbox{activation} & tanh & tanh & tanh\\
& \mbox{l2 regularization} & 0.01 & 0.01 & 0.01\\
& \mbox{batch normalization} & true & true & true\\
& \mbox{gaussian noise} & 0.12 & 0.12 & 0.12\\
& \mbox{dropout} & 0.01 & 0.01 & 0.01\\
\hline
\mbox{hidden2} & \mbox{neurons} & 1 & 7 & 7\\
& \mbox{activation} & tanh & tanh & tanh\\
& \mbox{l2 regularization} & 0.01 & 0.01 & 0.01\\
& \mbox{batch normalization} & true & true & true\\
& \mbox{gaussian noise} & 0.12 & 0.12 & 0.12\\
& \mbox{dropout} & 0.01 & 0.01 & 0.01\\
\hline
\mbox{output} & \mbox{neurons} & 1 & 2 (4) & 2(3) \\
& \mbox{activation} & exp & sigmoid & sigmoid\\
& \mbox{l2 regularization} & 0.01 & 0.01 & 0.01
\end{array}$
\caption{The specifications of the location, Weibull, and lognormal neural network, separated by layer.}
\label{tab:netParams}
\end{table}

\section{ProbSAT}
\label{sec:pSAT}
ProbSAT \cite{balint_choosing_2012} is a stochastic local search algorithm for solving the satisfiability problem.
It starts with a randomly generated initial assignment.
If the current assignment does not satisfy the formula, then ProbSAT randomly chooses an unsatisfied clause $C=(v_1 \lor v_2 \lor \dots)$.
Changing the value of the assignment for $v_i$ causes $C$ to be satisfied, this is called a flip (of $v_i$).  
ProbSAT flips $v_i$ with probability $\frac{f(v_i)}{\sum_{k=1}^{l} f(v_k)}$, where $f(x) = \frac{\textnormal{make}(x)^{c_m}}{(1+\textnormal{break}(x))^{c_b}}$.
The make value is the number of clauses which become satisfied if $x$ is flipped, while the break value is the number of clauses which become unsatisfied if 
$x$ is flipped. After a variable is chosen and flipped the new assignment is checked. If the new assignment still does not satisfy the formula, then
a new variable is chosen from another unsatisfied clause. This step is repeated until either a satisfying assignment is found or until the number of flips
exceed a certain $max\_flips$ value. If the $max\_flips$ value is exceeded, then ProbSAT is restarted with a new random assignment.

Note, that the function $f$ contains two parameters $c_b$ and $c_m$. Balint and Schöning \cite{balint_choosing_2012} found that $c_b = 2.3$ and $c_m=0$ are good choices for random $3-$SAT formulas close to the phase transition. These parameter settings are also used in this work.

\end{document}